\documentclass[twocolumn,pra,aps,floatfix]{revtex4}
\usepackage{amssymb}
\usepackage{amsfonts}
\usepackage{amsmath}
\usepackage{graphicx}

\begin{document}

\title{Quantifying environment non-classicality in dissipative open quantum
dynamics}
\date{\today }
\author{Adri\'{a}n A. Budini}
\affiliation{Consejo Nacional de Investigaciones Cient\'{\i}ficas y T\'{e}cnicas
(CONICET), Centro At\'{o}mico Bariloche, Avenida E. Bustillo Km 9.5, (8400)
Bariloche, Argentina, and Universidad Tecnol\'{o}gica Nacional (UTN-FRBA),
Fanny Newbery 111, (8400) Bariloche, Argentina}

\begin{abstract}
Open quantum systems are inherently coupled to their environments, which in
turn also obey quantum dynamical rules. By restricting to dissipative
dynamics, here we propose a measure that quantifies how far the environment
action on a system departs from the influence of classical noise
fluctuations. It relies on the lack of commutativity between the initial
reservoir state and the system-environment total Hamiltonian. Independently
of the nature of the dissipative system evolution, Markovian or
non-Markovian, the measure can be written in terms of the dual propagator
that defines the evolution of system operators. The physical meaning and
properties of the proposed definition are discussed in detail and also
characterized through different paradigmatic dissipative Markovian and
non-Markovian open quantum dynamics.
\end{abstract}

\maketitle

\section{Introduction}

Open quantum systems are inherently coupled to their supporting environments~%
\cite{breuerbook,vega}. This interaction induces time-irreversible behaviors
such as dissipation and decoherence. These phenomena have been studied in a
broad class of systems such as for example in quantum optics~\cite%
{carmichaelbook}, magnetic resonance~\cite{anderson,abragan}, solid state
devices~\cite{efenoise}, and quantum sensing~\cite{sensing}.

In a full microscopic description not only\ the system but also the
environment obeys quantum dynamical rules. Nevertheless, depending on system
and environment properties, as well as on the studied regimes, the system
fluctuations induced by the environment influence can be well approximated
by the action of classical stochastic fields. In fact, open quantum systems
driven by classical noises is a well established physical modelling~\cite%
{anderson,abragan,efenoise,sensing} that has been characterized from
different perspectives. Many specific studies rely on assuming, for example,
Gaussian~\cite{GaussianNoise,Gauss,morgado,cialdi,kelly} or telegraphic
noises~\cite{dico,dicoAbel,song}.

In the context of open quantum system theory it is of interest to
establishing the conditions under which the environment action can be
approximated by classical noises. For example, the possibility of
representing the system evolution in terms of a statistical superposition of
unitary dynamics has been explored recently~\cite{nori,Chen,franco,ChenChen}%
. When the open system dynamics only leads to dephasing~\cite%
{strunzTwoQubits,viola,roszak,kata,Cywinski,experK,clerck}, the possibility
of detecting quantum entanglement between the system and the environment~%
\cite{roszak,kata,Cywinski,experK} gives a solid criteria for determining
when a classical noise representation is appropriate or not. In addition
many related contributions focused on this problem, providing general or
particular conditions under which a classical representation is a valid
approximation~\cite%
{depol,spinbathnoise,shortParis,parislletA,bordone,hala,fabri,liu,Wang,fata,lika,Szanko,sun,liuBis}%
.

The main goal of this work is to introduce a measure that quantifies how
much the environment action departs from the influence of classical noise
fluctuations. This result provides an interesting insight and contribution
in the described research line. In contrast to previous analysis (see Refs.~%
\cite%
{nori,Chen,franco,ChenChen,strunzTwoQubits,viola,roszak,kata,Cywinski,experK}%
), here we are mainly interested in dissipative open quantum dynamics, that
is, the environment not only induce decoherence but also is able to induce
(energy) transitions between the system states.

The proposed measure has a clear physical motivation related to the
quantumness of the environment state and its dynamics. In addition, it is
valid independently of the system dynamical regime, that is, Markovian or
non-Markovian. In fact, independently of which approach is used to define
memory effects, operational~\cite{modi,budiniCPF} or non-operational~\cite%
{BreuerReview,plenioReview}, the proposed measure can be written in terms of
the dual evolution associated to system operators. Thus, it can be defined
consistently for Markovian Lindblad equations~\cite{alicki}\ but also, in
the same way, outside this regime. The proposal is characterized in detail
through its general properties and also through specific dissipative
Markovian and non-Markovian dynamics.

The manuscript is outlined as follows. In Sec. II we motivate and formulate
the environment non-classicality measure. In Sec. III it is characterized
for some general classes of open quantum dynamics. In Sec. IV we study its
behavior for specific Markovian and non-Markovian dissipative open system
dynamics. In Sec. V we provide the Conclusions.

\section{Measure of environment non-classicality}

In this section we introduce the environment non-classicality measure,
providing in addition some of its general properties.

\subsection{Physical motivation and definition}

We consider a system $(s)$ that interacts with its environment $(e).$ Their
quantum dynamics is set by a total Hamiltonian $H=H_{s}+H_{e}+H_{I},$ where $%
H_{s}$ and $H_{e}$ are the system and environment Hamiltonians respectively,
while $H_{I}$ defines their mutual interaction. The system density matrix $%
\rho _{t}$\ can be written as%
\begin{equation}
\rho _{t}=\mathbb{G}_{t,0}[\rho _{0}]\equiv \mathrm{Tr}_{e}[e^{-iHt}(\rho
_{0}\otimes \sigma _{0})e^{+iHt}].  \label{RhoUni}
\end{equation}%
Here, $\mathrm{Tr}[\cdots ]$ is the trace operation. Furthermore, we assume
uncorrelated $s$-$e$ initial conditions, where $\rho _{0}$ and $\sigma _{0}$
are the initial system and environment states respectively.

The quantum nature of system and the environment can be read
straightforwardly from Eq.~(\ref{RhoUni}). In fact, all objects appearing in
this expression can be written as matrixes that in general do not commutate.
Focussing on the environment, we argue that the nature of its influence over
the system is inherently quantum because in general its initial state does
not commutates with the total Hamiltonian, that is, $[H,\sigma _{0}]\neq 0.$
Supporting this argument, when the initial environment state approaches the
identity matrix $\sigma _{0}\simeq \mathrm{I}_{e},$ which implies $[H,\sigma
_{0}]\approx 0,$ its action over the system can be represented by classical
noises. This result is well known in the context of magnetic resonance~\cite%
{abragan} and also has been characterized when expressing the system
evolution in terms of stochastic wave vectors~\cite{SWF}. For systems
coupled to thermal environments the property $[H,\sigma _{0}]\approx 0$
becomes valid in a high temperature limit.

Under the motivation of the previous perspective, we rewrite the system
state [Eq.~(\ref{RhoUni})] as%
\begin{equation}
\rho _{t}=\mathrm{Tr}_{e}[(e^{-iHt}\rho _{0}e^{+iHt})\sigma _{0}]+\mathrm{Tr}%
_{e}[(e^{-iHt}\rho _{0}e^{+iHt})\Delta \sigma _{t}],  \label{RhoSplit}
\end{equation}%
where $\Delta \sigma _{t}\equiv e^{-iHt}\sigma _{0}e^{+iHt}-\sigma _{0}.$ We
identify the first term with the \textquotedblleft
classical\textquotedblright\ contribution of the environment influence. In
fact, when $[H,\sigma _{0}]\approx 0$ the first contribution does not vanish
while the second one fade out (vanishes) correspondingly. Interestingly,
when describing the system evolution in a weak interaction and Markovian
limits the previous splitting recovers the structure of quantum master
equations~\cite{abragan,goldman,ZM} proposed for dealing with a high
temperature approximation~\cite{abragan}.

While the property $\mathrm{Tr}_{s}[\rho _{t}]=1$ is fulfilled, each
contribution in Eq.~(\ref{RhoSplit}) does not preserve trace of the system
by itself. Thus, for measuring departure of the environment action with
respect to classical noises we arrive to the (dimensionless) time-dependent
quantumness measure%
\begin{equation}
Q_{t}\equiv \mathrm{Tr}_{se}[(e^{-iHt}\rho _{0}e^{+iHt})\sigma _{0}].
\label{QDef}
\end{equation}%
It corresponds to the trace over the system degrees of freedom of the first
term in the splitting~(\ref{RhoSplit}).

\subsection{Degree of environment quantumness}

The definition~(\ref{QDef}) has some desirables properties. For example,
when $[H,\sigma _{0}]=0$ it follows $Q_{t}=1.$ Therefore, this value
indicates classicality. On the other hand, it is straightforward to obtain%
\begin{equation}
\frac{dQ_{t}}{dt}=-i\mathrm{Tr}_{se}[(e^{-iHt}\rho _{0}e^{+iHt})[H,\sigma
_{0}]],
\end{equation}%
while its $n$-time-derivative reads%
\begin{equation}
\frac{d^{n}Q_{t}}{dt^{n}}=(-i)^{n}\mathrm{Tr}_{se}[(e^{-iHt}\rho
_{0}e^{+iHt})[H^{(n)},\sigma _{0}]],
\end{equation}%
where $[H^{(1)},\sigma _{0}]=[H,\sigma _{0}],\ [H^{(2)},\sigma
_{0}]=[H,[H,\sigma _{0}]],$ and in general $[H^{(n)},\sigma
_{0}]=[H,[H^{(n-1)},\sigma _{0}].$ From these expressions we conclude that
the time-derivatives of $Q_{t}$ are proportional to the lack of
commutativity of $\sigma _{0}$ with higher nested commutators of the total
Hamiltonian $H.$

In spite of the previous properties, the definition of $Q_{t}$ [Eq.~(\ref%
{QDef})] is symmetrical in the initial system and environment states. In
particular, when $\rho _{0}=\mathrm{I}_{s}/\dim (\mathcal{H}_{s}),$ where $%
\mathrm{I}_{s}$ is the identity matrix and $\dim (\mathcal{H}_{s})$\ is the
dimension of the system Hilbert space, it follows that $Q_{t}=1.$

The difference between the roles played by the system and the environment is
introduced by defining a \textquotedblleft degree of environment
quantumness,\textquotedblright\ denoted as $D_{Q},$ which reads%
\begin{equation}
D_{Q}\equiv \max_{\lbrack \rho _{0}]}\left\vert \lim_{t\rightarrow \infty
}\int_{0}^{t}dt^{\prime }\frac{dQ(t^{\prime })}{dt^{\prime }}\right\vert
=\max_{[\rho _{0}]}\left\vert \lim_{t\rightarrow \infty }Q_{t}-1\right\vert .
\label{Degree}
\end{equation}%
Here, it was used that $Q_{0}=1.$ Furthermore, it is assumed that a
stationary regime is achieved. The maximization is over the initial system
state $\rho _{0}.$ The equality $D_{Q}=0$ is then associated with
environment classicality. In addition, this parameter allows to study $Q_{t}$%
\ [Eq.~(\ref{QDef})] by choosing system initial conditions that maximize $%
D_{Q}.$

\subsection{Definition in terms of the operator dual evolution}

Given a system operator $A,$ by definition its expectation value reads $%
\langle A\rangle _{t}\equiv \mathrm{Tr}_{s}[\rho _{t}A].$ It can
alternatively be written as $\langle A\rangle _{t}=\mathrm{Tr}_{s}[\rho
_{0}A_{t}]=\mathrm{Tr}_{s}[\rho _{0}\mathbb{G}_{t,0}^{\bigstar }[A_{0}]]$
where the dual propagator $\mathbb{G}_{t,0}^{\bigstar }$ for the operator
evolution, from Eq.~(\ref{RhoUni}), is given by%
\begin{equation}
A_{t}=\mathbb{G}_{t,0}^{\bigstar }[A_{0}]\equiv \mathrm{Tr}%
_{e}[e^{+iHt}Ae^{-iHt}\sigma _{0}].  \label{AtDefinition}
\end{equation}%
Therefore, from this expression and Eq.~(\ref{QDef}) it follows that the
quantumness measure $Q_{t}$ can be written as%
\begin{equation}
Q_{t}=\mathrm{Tr}_{s}[\mathbb{G}_{-t,0}^{\bigstar }[\rho _{0}]],
\label{QDual}
\end{equation}%
which only depends on the operator dual evolution and the initial system
state. From this expression it follows that $Q_{t}/\dim (\mathcal{H}_{s})$
can be read as the expectation value of the \textquotedblleft
operator\textquotedblright\ $\rho _{0}$ at time $t$ given that the
\textquotedblleft initial system state\textquotedblright\ is $\mathrm{I}%
_{s}/\dim (\mathcal{H}_{s})$ [see Eqs.~(\ref{AtDefinition}) and (\ref{QDual}%
)].

Taking into account that $\rho _{0}$ is a positive definite operator, this
equivalent interpretation of $Q_{t}$ allow us to obtain%
\begin{equation}
0\leq Q_{t}\leq \dim (\mathcal{H}_{s}).  \label{Bounds}
\end{equation}%
This inequality is valid independently of the system and environment initial
conditions and also of the particular microscopic model. On the other hand,
it consistently implies that the quantumness of the environment influence is
bounded by the dimension of the system Hilbert space. In addition, Eq.~(\ref%
{Bounds}) implies%
\begin{equation}
0\leq D_{Q}\leq \dim (\mathcal{H}_{s})-1,  \label{DqBounds}
\end{equation}%
where classicality corresponds to $D_{Q}=0.$

\subsection{Optimal initial states}

By analyzing the stationary regime of Eq.~(\ref{QDef}), or alternatively
Eq.~(\ref{QDual}), it follows that%
\begin{equation}
\lim_{t\rightarrow \infty }Q_{t}=\dim (\mathcal{H}_{s})\mathrm{Tr}_{s}[%
\tilde{\rho}_{\infty }\rho _{0}],  \label{QinfExplicito}
\end{equation}%
where the stationary system state is $\tilde{\rho}_{\infty }\equiv
\lim_{t\rightarrow \infty }\tilde{\rho}_{t}.$ The upper tilde symbol
represents a time-reversal operation, $t\leftrightarrow -t.$ Eqs.~(\ref%
{Degree}) and~(\ref{QinfExplicito}) implies that%
\begin{equation}
\frac{D_{Q}}{\dim (\mathcal{H}_{s})}=\max_{[\rho _{0}]}\left\vert \mathrm{Tr}%
_{s}[\tilde{\rho}_{\infty }\rho _{0}]-\frac{1}{\dim (\mathcal{H}_{s})}%
\right\vert .  \label{DqInfinita}
\end{equation}%
Thus, $D_{Q}$ can be seen as a functional of $\rho _{0}$ that is
parametrized by the stationary state $\rho _{\infty }.$ The optimal state $%
\rho _{0}$ that maximizes $D_{Q}$ is obtained below.

The expression (\ref{DqInfinita}) provides a clear geometric interpretation
of $D_{Q}.$ Introducing a basis of vectors $\{|i\rangle \}$ where the
stationary system state is a diagonal matrix, $\tilde{\rho}_{\infty
}=\sum_{i}\lambda _{i}|i\rangle \langle i|,$ with $i=1,\cdots \dim (\mathcal{%
H}_{s}),$ it follows that $D_{Q}=\max_{\{p_{i}\}}\left\vert \dim (\mathcal{H}%
_{s})\sum_{i}\lambda _{i}p_{i}-1\right\vert ,$ where $p_{i}\equiv \langle
i|\rho _{0}|i\rangle .$ Therefore, $D_{Q}$ is the maximal (absolute) value
assumed by the hyperplane defined by the variables $\{p_{i}\}$ when
restricted to the domain $\sum_{i}p_{i}=1.$ It is simple to bound the main
contribution to $D_{Q}$ as $\sum_{i}\lambda _{i}p_{i}\leq
(\sum_{i}p_{i})\max (\{\lambda _{i}\})=\max (\{\lambda _{i}\}).$ This
boundary is always achieved by choosing $\rho _{0}$ as the eigenprojector of 
$\tilde{\rho}_{\infty }$\ with the maximal eigenvalue. Thus, we conclude that%
\begin{equation}
D_{Q}=\dim (\mathcal{H}_{s})\max (\{\lambda _{i}\})-1,\ \ \ \ \ \ \rho
_{0}=|i_{\max }\rangle \langle i_{\max }|,  \label{MaxEigen}
\end{equation}%
where $\max (\{\lambda _{i}\})$ is the largest eigenvalue of the stationary
state $\tilde{\rho}_{\infty }\equiv \lim_{t\rightarrow \infty }\tilde{\rho}%
_{t},$ while $|i_{\max }\rangle $ is the corresponding eigenstate, $\tilde{%
\rho}_{\infty }|i_{\max }\rangle =\max (\{\lambda _{i}\})|i_{\max }\rangle .$
This expression for $D_{Q}$ is valid when the stationary state does not
depends on the initial condition. In addition, we notice that in general
more than one initial state, $\rho _{0}\neq |i_{\max }\rangle \langle
i_{\max }|,$\ may lead to this extreme value (see Sec. IV). On the other
hand, it is simple to realize that when the time-reversal operation is
equivalent to conjugation Eq.~(\ref{MaxEigen}) is valid with $\tilde{\rho}%
_{\infty }\rightarrow \rho _{\infty }.$

\section{Classicality for different classes of open system dynamics}

In this Section we characterize the previous proposal for different classes
of open quantum system dynamics where the quantumness measure indicates
classicality, $Q_{t}=1.$

\subsection{Hamiltonian ensembles}

In Refs.~\cite{nori,Chen,franco,ChenChen} the classicality of the
environment action was related to the possibility of representing the open
system dynamics in terms of Hamiltonian ensembles, that is, a statistical
superposition of different system unitary dynamics. This kind of dynamics is
recovered in the present approach after assuming that $[H,\sigma _{0}]=0.$
In fact, introducing a complete basis of environment states $\{|e\rangle \}$
where the initial state is diagonal, $\sigma _{0}=\sum_{e}p_{e}|e\rangle
\langle e|,$ with $p_{e}=\langle e|\sigma _{0}|e\rangle ,$ the system
density matrix [Eq.~(\ref{RhoUni})] can be written as%
\begin{equation}
\rho _{t}=\sum_{e}p_{e}e^{-itH_{s}^{(e)}}\rho _{0}e^{+itH_{s}^{(e)}},\ \ \ \
\Rightarrow \ \ \ \ Q_{t}=1,  \label{HEnsemble}
\end{equation}%
where the system Hamiltonians are $H_{s}^{(e)}\equiv H_{s}+\langle
e|(H_{e}+H_{I})|e\rangle .$ The equality $Q_{t}=1$ follows from Eq.~(\ref%
{QDef}) and is valid independently of the system initial condition. The
degree of quantumness [Eq.~(\ref{Degree})] also indicates the presence of a
classical environment influence, $D_{Q}=0.$

\subsection{Stochastic Hamiltonians}

The coupling of a quantum system to classical noises is usually modelled by
(system) stochastic Hamiltonians $H_{st}(t).$ Their time-dependence take
into account the action of classical fluctuating external fields. The noises
can have arbitrary statistical properties (see for example Refs.~\cite%
{GaussianNoise,Gauss,morgado,cialdi,kelly,dico,dicoAbel,song}). Even more,
their correlation can also be arbitrary, that is, delta correlated (white
noises) or color ones (finite correlation times).

For each realization of the noises, we introduce the stochastic propagator $%
\mathcal{T}_{st}(t)=\lceil \exp -i\int_{0}^{t}dt^{\prime }H_{st}(t^{\prime
})\rceil ,$ where $\lceil \cdots \rceil $ means a time-ordering operation.
The system density matrix can then be written as%
\begin{equation}
\rho _{t}=\overline{\mathcal{T}_{st}(t)\rho _{0}\mathcal{T}_{st}^{\dagger
}(t)},\ \ \ \ \Rightarrow \ \ \ \ Q_{t}=1,  \label{Ruido}
\end{equation}%
where the overline denotes an average over noise realizations. The equality $%
Q_{t}=1$ is valid for arbitrary system initial conditions. It can be derived
by using the alternative definition in terms of the dual evolution [Eq.~(\ref%
{QDual})]. Consistently, in this case $D_{Q}=0.$

\subsection{Collisional dynamics}

Open quantum systems dynamics, in Markovian and non-Markovian regimes, can
also be modelled through collisional models~\cite%
{collisional,colisionVacchini}. The underlying stochastic dynamics consists
in free propagation with the system Hamiltonian added to the action of an
instantaneous transformation that occurs at successive random times. The
(stochastic) state of the system conditioned to the occurrence of $n$%
-collisional-events can be written as%
\begin{equation}
\rho _{t}^{(n)}=\mathcal{G}_{t-t_{n}}\mathcal{EG}_{t_{n}-t_{n-1}}\cdots 
\mathcal{EG}_{t_{2}-t_{1}}\mathcal{EG}_{t_{1}}[\rho _{0}].
\end{equation}%
Here, $\mathcal{G}_{t}$ is the propagator of the free evolution, $\mathcal{G}%
_{t}[\bullet ]\equiv \exp [-itH_{s}]\bullet \exp [+itH_{s}],$ while $%
\mathcal{E}$ is an arbitrary completely positive trace preserving
superoperator. The times $\{t_{i}\}_{i=1}^{n}$ are random variables in the
interval $(0,t).$ The state of the system follows as%
\begin{equation}
\rho _{t}=\sum_{n=0}^{\infty }\overline{\rho _{t}^{(n)}},  \label{Colision}
\end{equation}%
where the overline means an average over the random collisional times. The
dual operator dynamics can be written in a similar way. In the Appendix we
develop a formal derivation.

Using the definition~(\ref{QDual}) it is possible to conclude (see Appendix)
that%
\begin{equation}
\mathrm{Tr}_{s}[\mathcal{E}^{\bigstar }[A]]=\mathrm{Tr}_{s}[A]\Rightarrow \
\ \ \ \ Q_{t}=1,  \label{UnitalCollision}
\end{equation}%
where the dual superoperator is defined from the relation $\mathrm{Tr}_{s}[A%
\mathcal{E}[\rho ]]=\mathrm{Tr}_{s}[\rho \mathcal{E}^{\bigstar }[A]],$ with $%
A$ being an arbitrary system operator. Thus, when the dual superoperator $%
\mathcal{E}^{\bigstar }$ preserves trace the quantumness indicator vanishes
identically for any system initial condition, which in turn implies $%
D_{Q}=0. $ The condition~(\ref{UnitalCollision}) is fulfilled when $\mathcal{%
E}$ corresponds to a unitary transformation and in general is fulfilled by
unital maps (see below).

\subsection{Lindblad equations}

When the system-environment coupling is weak and the time-correlations of
environment operators define the minor time-scale of the problem, a
Born-Markov approximation applies. Discarding non-secular terms, the system
evolution can be written as a Lindblad equation~\cite{breuerbook,alicki},%
\begin{equation}
\frac{d\rho _{t}}{dt}=-i[\bar{H}_{,}\rho _{t}]+\sum_{\mu \nu }a_{\mu \nu
}(V_{\mu }\rho _{t}V_{\nu }^{\dagger }-\frac{1}{2}\{V_{\nu }^{\dagger
}V_{\mu },\rho _{t}\}_{+}).  \label{LindbladGen}
\end{equation}%
Here, $\bar{H}$ is an effective system Hamiltonian that may include
contributions induced by the interaction with the environment. $\{V_{\mu }\}$
are system operators, while the matrix of rate coefficients $\{a_{\mu \nu
}\} $ defines a semi-positive definite matrix. The anticommutator operation
is defined as $\{a,c\}_{+}\equiv (ac+ca).$

The quantumness measure $Q_{t}$ can be calculated for the previous quantum
master equation by using its definition in terms of the dual evolution [Eq.~(%
\ref{QDual})]. For an arbitrary operator $A_{t}$ it reads~\cite{alicki}%
\begin{equation}
\frac{dA_{t}}{dt}=+i[\bar{H},A_{t}]+\sum_{\mu \nu }a_{\mu \nu }(V_{\nu
}^{\dagger }A_{t}V_{\mu }-\frac{1}{2}\{V_{\nu }^{\dagger }V_{\mu
},A_{t}\}_{+}).  \label{At}
\end{equation}%
Consistently, notice that this evolution does not preserve trace. By solving
Eq.~(\ref{At}) with initial condition $A_{t}|_{t=0}=\rho _{0},$ the
quantumness measure can be written as $Q_{t}=\mathrm{Tr}_{s}[\tilde{A}_{t}],$
where the tilde symbol takes into account the time reversal operation $%
t\leftrightarrow -t.$ In this way, the present approach can be applied in
the Markovian regime where a Lindblad equation approximate the open system
dynamics.

Interestingly, Lindblad equations that are compatible with the influence of
classical noises have been characterized from a rigorous mathematical point
of view~\cite{viola,mathBiStoch}. The proposed structures were derived as
commutative dilations of dynamical semigroups. Specifically, they correspond
to Eq.~(\ref{LindbladGen}) written in a diagonal base of operators $(a_{\mu
\nu }=\delta _{\mu \nu }a_{\mu })$ with the constraints of Hermitian
operators, $V_{\mu }=V_{\mu }^{\dagger },$ or alternatively unitary ones, $%
V_{\mu }^{\dagger }V_{\mu }=\mathrm{I}_{s}.$ These solutions can be put
in-one-to-one correspondence with models based respectively on stochastic
Hamiltonians [Eq.~(\ref{Ruido})] with white-noise fluctuations and
collisional models [Eq.~(\ref{Colision}) and~(\ref{UnitalCollision})] with
Poisson statistics between collisional events.

\subsection{Unital open system dynamics}

A completely positive open system dynamics can always be written in a Krauss
representation as~\cite{breuerbook}%
\begin{equation}
\rho _{t}=\sum_{\alpha }T_{\alpha }\rho _{0}T_{\alpha }^{\dagger },\ \ \ \ \
\ \ \ \ \ \sum_{\alpha }T_{\alpha }^{\dagger }T_{\alpha }=\mathrm{I}_{s},
\end{equation}%
where the system operators $\{T_{\alpha }\}$ are time-dependent, $T_{\alpha
}=T_{\alpha }(t).$ The dynamics is defined as unital when in addition it is
fulfilled that%
\begin{equation}
\sum_{\alpha }T_{\alpha }T_{\alpha }^{\dagger }=\mathrm{I}_{s},\ \ \ \
\Rightarrow \ \ \ \ Q_{t}=1.  \label{Unital}
\end{equation}%
We notice that the result $Q_{t}=1,$ valid for arbitrary system initial
conditions, follows from Eq.~(\ref{QDual}) and after noting that the dual
operator evolution can be written as $A_{t}=\sum_{\alpha }T_{\alpha
}^{\dagger }A_{0}T_{\alpha },$ which implies that $\mathrm{Tr}%
_{s}[A_{t}]=\sum_{\alpha }\mathrm{Tr}_{s}[T_{\alpha }T_{\alpha }^{\dagger
}A_{0}]=\mathrm{Tr}_{s}[A_{0}].$

In general it is possible to argue that any open quantum dynamics induced by
coupling the system with stochastic classical degrees of freedom is always
unital, which consistently implies $Q_{t}=1.$ In fact, the dynamics defined
by Eqs.~(\ref{Ruido}) and~(\ref{Colision}) can be read as \textquotedblleft
non-Markovian\textquotedblright\ extensions of the commutative dilations of
dynamical semigroups obtained in Ref.~\cite{mathBiStoch}. With non-Markovian
here we mean considering non-white noises or non-Poisson statistics
respectively.

On the other hand, the inverse implication is not valid in general, that is,
there exist unital dynamics that cannot be obtained by considering the
action of classical stochastic fields. This property emerges, for example,
in dephasing dynamics with $\dim (\mathcal{H}_{s})\geq 3$~\cite{viola}. In
addition, this feature has been related to the break of a time-reversal
symmetry~\cite{clerck}. While these cases implies a limitation on the
applicability of the indicator $Q_{t},$ the corresponding class of dynamics
is well characterized. On the other hand, the examples studied in the next
section explicitly demonstrate the consistence of the proposed approach.

\section{Examples}

Here the quantumness indicator $Q_{t}$ is characterized for some specific
dissipative Markovian and non-Markovian open quantum dynamics.

\subsection{Two-level system in contact with a thermal environment}

We consider a two-level system interacting with a Bosonic bath at
temperature $T.$ It density matrix $\rho _{t}$ evolves as~\cite{breuerbook}%
\begin{eqnarray}
\frac{d\rho _{t}}{dt} &=&\frac{-i\omega _{0}}{2}[\sigma _{z},\rho
_{t}]+\kappa (\sigma \rho _{t}\sigma ^{\dagger }-\frac{1}{2}\{\sigma
^{\dagger }\sigma ,\rho _{t}\}_{+})  \notag \\
&&\ \ \ \ \ \ \ \ \ \ \ \ \ \ \ \ \ +\zeta (\sigma ^{\dagger }\rho
_{t}\sigma -\frac{1}{2}\{\sigma \sigma ^{\dagger },\rho _{t}\}_{+}).
\label{Thermal}
\end{eqnarray}%
With $\sigma _{z}$ we denote the $z$-Pauli matrix. $\sigma $ and $\sigma
^{\dagger }$ are the standard lowering and raising operators with respect to
the eigenvectors of $\sigma _{z}.$ Furthermore, $\kappa =\gamma (n_{th}+1)$
and $\zeta =\gamma n_{th},$ where $\gamma $\ is the natural decay rate and $%
n_{th}=\exp (-\beta \hbar \omega _{0})/[1-\exp (-\beta \hbar \omega _{0})]$
is the average number of thermal boson excitations at the natural frequency
of the system, with $\beta =1/kT.$

Using the alternative definition~(\ref{QDual}), jointly with the dual
evolution~(\ref{At}), for arbitrary system initial conditions it is possible
to obtain%
\begin{equation}
Q_{t}=1+\langle \sigma _{z}\rangle _{\infty }\langle \sigma _{z}\rangle
_{0}[1-e^{-t(\kappa +\zeta )}],  \label{QThermico}
\end{equation}%
where the operator mean values are $\langle \sigma _{z}\rangle _{0}=\mathrm{%
Tr}_{s}[\sigma _{z}\rho _{0}]$ and $\langle \sigma _{z}\rangle _{\infty
}=\lim_{t\rightarrow \infty }\mathrm{Tr}_{s}[\sigma _{z}\rho _{t}]=(\zeta
-\kappa )/(\zeta +\kappa )\leq 0.$ In general, depending on the initial
condition, as a function of time $Q_{t}$ decays or grows in a monotonic way.
In any of these cases, consistently with Eq.~(\ref{Bounds}), it is fulfilled
that $0\leq Q_{t}\leq 2.$

From Eq.~(\ref{QThermico}) it follows that $\lim_{t\rightarrow \infty
}Q_{t}=1+\langle \sigma _{z}\rangle _{\infty }\langle \sigma _{z}\rangle
_{0}.$ This stationary value has maximal departure from the unity value when 
$\langle \sigma _{z}\rangle _{0}=\pm 1.$ Thus, the initial conditions that
maximize the definition of $D_{Q}$ [Eq.~(\ref{Degree})] are pure states,
which in turn are eigenvectors of $\sigma _{z}.$ This result is consistent
with Eq.~(\ref{MaxEigen}). The degree of environment quantumness finally
reads%
\begin{equation}
D_{Q}=|\langle \sigma _{z}\rangle _{\infty }|=\left\vert \frac{\zeta -\kappa 
}{\zeta +\kappa }\right\vert =\tanh \Big{(}\beta \frac{\hbar \omega _{0}}{2}%
\Big{)}.  \label{DqThermal}
\end{equation}%
In the last equality we have used the dependence on temperature of the
characteristic rates.

Eq.~(\ref{DqThermal}) defines the degree of environment quantumness
corresponding to the evolution~(\ref{Thermal}). As a function of the inverse
temperature $\beta $\ it has the expected behaviors. In fact, in the limit
of high temperatures it follows $\lim_{\beta \rightarrow 0}D_{Q}=0,$ which
correctly means that the environment influence can be represented through
classical noises~\cite{abragan,SWF}. In the limit of vanishing temperatures $%
D_{Q}$ assumes its maximal value [Eq.~(\ref{DqBounds})], $\lim_{\beta
\rightarrow \infty }D_{Q}=1.$

\subsection{Non-Markovian decay at zero temperature}

In contrast to the previous case, here we consider a dynamics where the
Born-Markov approximation does not applies in general. The microscopic
dynamics is defined by the Hamiltonians $H_{s}=(\omega _{0}/2)\sigma _{z},$ $%
H_{e}=\sum_{j}\omega _{k}a_{k}^{\dagger }a_{k},$ while the interaction is
sets by $H_{I}=\sum_{k}(g_{k}\sigma ^{\dagger }a_{k}+g_{k}^{\ast }\sigma
a_{k}^{\dagger }).$ With $a_{k}$ and $a_{k}^{\dagger }$ we denote the
annihilation and creation operators associated to each mode of the Bosonic
environment. Memory effects for this open dynamics has been studied from
both non-operational~\cite{breuerDecayTLS} and operational~\cite%
{budiniBrasil} approaches to quantum non-Markovianity

The (two-level) system dynamics can be solved in an exact way by assuming
that all modes of the environment begin in their ground states, which is
equivalent to a vanishing temperature assumption. The system density matrix
reads~\cite{breuerbook}%
\begin{equation}
\rho _{t}=\left( 
\begin{array}{cc}
\rho _{0}^{++}|c_{t}|^{2} & \rho _{0}^{+-}\ c_{t} \\ 
\rho _{0}^{-+}\ c_{t}^{\ast } & \ \ \ \rho _{0}^{--}+\rho
_{0}^{++}(1-|c_{t}|^{2})%
\end{array}%
\right) .  \label{RhoExacta}
\end{equation}%
Here, $\rho _{0}^{ss^{\prime }}\equiv \langle s|\rho _{0}|s^{\prime }\rangle
,$ where $\{|s\rangle \}=|{\pm \rangle }$ are the eigenvectors of $\sigma
_{z}.$ The function $c_{t}$ is defined by $(d/dt)c(t)=-\int_{0}^{t}f(t-t^{%
\prime })c(t^{\prime })dt^{\prime },$ where the memory kernel corresponds to
the bath correlation function $f(t)\equiv \sum_{k}|g_{k}|^{2}\exp [+i(\omega
_{0}-\omega _{k})t].$

Using that $\langle A\rangle _{t}=\mathrm{Tr}_{s}[\rho _{t}A]=\mathrm{Tr}%
_{s}[\rho _{0}A_{t}],$ from Eq.~(\ref{RhoExacta}) it is possible to obtain
the operator dual dynamics, which explicitly reads%
\begin{equation}
A_{t}=\left( 
\begin{array}{cc}
A_{0}^{++}|c_{t}|^{2}+A_{0}^{--}(1-|c_{t}|^{2})\ \ \  & A_{0}^{+-}\
c_{t}^{\ast } \\ 
A_{0}^{-+}\ c_{t} & A_{0}^{--}%
\end{array}%
\right) ,
\end{equation}%
where $A_{0}^{ss^{\prime }}\equiv \langle s|A_{0}|s^{\prime }\rangle .$ The
environment quantumness $Q_{t}$ can be obtained from the relation~(\ref%
{QDual}), which here delivers%
\begin{equation}
Q_{t}=1-\langle \sigma _{z}\rangle _{0}[1-|c_{t}|^{2}],  \label{QExact}
\end{equation}%
where $\langle \sigma _{z}\rangle _{0}=\mathrm{Tr}_{s}[\sigma _{z}\rho
_{0}]. $

From Eq.~(\ref{QExact}) it follows that\ $\lim {}_{t\rightarrow \infty
}Q_{t}=1-\langle \sigma _{z}\rangle _{0}.$ This limit assumes extreme values
when $\langle \sigma _{z}\rangle _{0}=\pm 1.$ Therefore, the degree of
environment quantumness is maximal [Eq.~(\ref{DqBounds})],%
\begin{equation}
D_{Q}=1.  \label{DqUnoNoMarkov}
\end{equation}%
Consistently, this value also emerges from the Lindblad modeling Eq.~(\ref%
{Thermal}) when the environment temperature vanishes [see Eq.~(\ref%
{DqThermal})]. In addition, the expressions for $Q_{t},$ Eq.~(\ref{QThermico}%
) with $\langle \sigma _{z}\rangle _{\infty }=-1$ and Eq.~(\ref{QExact}),
assume the same structure. The non-Markovian effects appear through the time
behavior of the decay function $|c_{t}|^{2},$ which in contrast to the
Markovian case, may develop oscillatory behaviors~\cite{breuerbook}. For
example, assuming a Lorentzian spectral density, which implies the
exponential correlation $f(t)=(\gamma /2\tau _{c})\exp [-|t|/\tau _{c}],$ it
follows $c_{t}=e^{-t/2\tau _{c}}[\cosh (t\chi /2\tau _{c})+\chi ^{-1}\sinh
(t\chi /2\tau _{c})],$ where $\chi \equiv \sqrt{1-2\gamma \tau _{c}}.$ In a
weak coupling limit $\gamma \ll 1/\tau _{c},$ where the correlation time $%
\tau _{c}$\ of the bath is the minor time scale of the problem, an monotonic
exponential decay is recovered $c_{t}\simeq \exp (-\gamma t/2)$ [Eq.~(\ref%
{QThermico})].

\subsection{Resonance fluorescence}

An optical two-level transition submitted to the action of a resonant
external laser field can be well approximated through the evolution~\cite%
{carmichaelbook}%
\begin{equation}
\frac{d\rho _{t}}{dt}=-i\frac{\Omega }{2}[\sigma _{x},\rho _{t}]+\gamma
(\sigma \rho _{t}\sigma ^{\dagger }-\frac{1}{2}\{\sigma ^{\dagger }\sigma
,\rho _{t}\}_{+}).  \label{Fluor}
\end{equation}%
Here, $\gamma $ is the natural decay rate while the frequency $\Omega $ is
proportional to the intensity of the external excitation. With $\sigma _{j}$ 
$(j=x,y,z)$ we denote the $j$-Pauli matrix. As before, $\sigma $ and $\sigma
^{\dagger }$ are the standard lowering and raising operators. We notice that
the effective environment action correspond to a thermal bath a cero
temperature. Below we study how the previous result $D_{Q}=1$ [Eqs.~(\ref%
{DqThermal})] is affected by the presence of the external excitation.

From Eqs.~(\ref{QDual}) and (\ref{At}), in a Laplace domain $%
[f(u)=\int_{0}^{\infty }dte^{-ut}f(t)],$\ $Q_{t}$ is defined by the exact
expression%
\begin{eqnarray}
Q_{u} &=&\frac{1}{u}-\langle \sigma _{z}\rangle _{0}\frac{\gamma (2u+\gamma )%
}{u[(u+\gamma )(2u+\gamma )+2\Omega ^{2}]}  \notag \\
&&+\langle \sigma _{y}\rangle _{0}\frac{2\gamma \Omega }{u[(u+\gamma
)(2u+\gamma )+2\Omega ^{2}]},  \label{QLaplace}
\end{eqnarray}%
where $\langle \sigma _{j}\rangle _{0}=\mathrm{Tr}_{s}[\sigma _{j}\rho
_{0}]. $ Explicitly,%
\begin{equation}
\langle \sigma _{z}\rangle _{0}=(\rho _{0}^{++}-\rho _{0}^{--}),\ \ \ \ \ \
\langle \sigma _{y}\rangle _{0}=i(\rho _{0}^{+-}-\rho _{0}^{-+}).
\end{equation}%
In the time domain, from Eq.~(\ref{QLaplace}) it is possible to write%
\begin{equation}
Q_{t}=1+\int_{0}^{t}\gamma dt^{\prime }e^{-\frac{3}{4}\gamma t^{\prime
}}[\langle \sigma _{z}\rangle _{0}\ z(t^{\prime })+\langle \sigma
_{y}\rangle _{0}\ y(t^{\prime })],  \label{QdeTe}
\end{equation}%
where the auxiliary functions are 
\begin{subequations}
\begin{eqnarray}
y(t) &\equiv &4\frac{\Omega }{\Gamma }\sinh \Big{(}\frac{t\Gamma }{4}\Big{)},
\\
z(t) &\equiv &\cosh \Big{(}\frac{t\Gamma }{4}\Big{)}-\frac{\gamma }{\Gamma }%
\sinh \Big{(}\frac{t\Gamma }{4}\Big{)},
\end{eqnarray}%
with $\Gamma \equiv \sqrt{\gamma ^{2}-(4\Delta )^{2}}.$

The stationary value of $Q_{t}$ can be obtained straightforwardly from Eq.~(%
\ref{QLaplace}) as $Q_{\infty }=\lim_{t\rightarrow \infty
}Q_{t}=\lim_{u\rightarrow 0}uQ_{u},$ which leads to 
\end{subequations}
\begin{equation}
Q_{\infty }=1+\langle \sigma _{z}\rangle _{\infty }\langle \sigma
_{z}\rangle _{0}+\langle \sigma _{y}\rangle _{\infty }\langle \sigma
_{y}\rangle _{0},  \label{QInfinito}
\end{equation}%
where the stationary mean values are%
\begin{equation}
\langle \sigma _{z}\rangle _{\infty }=-\frac{\gamma ^{2}}{\gamma
^{2}+2\Omega ^{2}},\ \ \ \ \ \langle \sigma _{y}\rangle _{\infty }=\frac{%
2\gamma \Omega }{\gamma ^{2}+2\Omega ^{2}}.
\end{equation}%
Consistently, these expressions also follow as $\langle \sigma _{j}\rangle
_{\infty }=\lim_{t\rightarrow \infty }\mathrm{Tr}_{s}[\sigma _{j}\rho _{t}].$

The system initial conditions that lead to extreme values of $(Q_{\infty
}-1) $ can be determine from Eq.~(\ref{QInfinito}). It is found that the
initial state must be pure, $\rho _{0}=|\psi _{0}\rangle \langle \psi _{0}|,$
where the state $|\psi _{0}\rangle $ is parametrized in terms of angles $%
(\theta _{0},\phi _{0})$ in the Bloch sphere, $|\psi _{0}\rangle =|\psi (\pm
,\theta _{0},\phi _{0})\rangle $ \cite{shankar}. Maximization implies that 
\begin{subequations}
\label{Angles}
\begin{equation}
\tan (\theta _{0})=-\frac{2\Omega }{\gamma }=\frac{\langle \sigma
_{y}\rangle _{\infty }}{\langle \sigma _{z}\rangle _{\infty }},\ \ \ \ \ \ \
\ \ \ \phi _{0}=\frac{\pi }{2}.
\end{equation}%
We remark that there are two different orthogonal states $\{|\psi (\pm
,\theta _{0},\phi _{0})\rangle \}$ associated to the direction defined by
these angles. They correspond the basis where the stationary state $\rho
_{\infty }=\lim {}_{t\rightarrow \infty }\rho _{t},$ is a diagonal matrix.
In addition, it is found that maximization is achieved with%
\begin{equation}
\tan (\tilde{\theta}_{0})=\frac{2\Omega }{\gamma }=-\frac{\langle \sigma
_{y}\rangle _{\infty }}{\langle \sigma _{z}\rangle _{\infty }},\ \ \ \ \ \ \
\ \ \ \tilde{\phi}_{0}=\frac{3\pi }{2}.
\end{equation}%
These angles define the basis where the (time-reversed) stationary system
density matrix is a diagonal operator, $\tilde{\rho}_{\infty }=\lim
{}_{t\rightarrow \infty }\rho _{t}^{\ast },$ where conjugation is taken in
the basis defined by the eigenvectors of $\sigma _{z}.$ These last solutions
are consistent with Eq.~(\ref{MaxEigen}). 
\begin{figure}[tbp]
\includegraphics[bb=46 870 730 1132,angle=0,width=8.85cm]{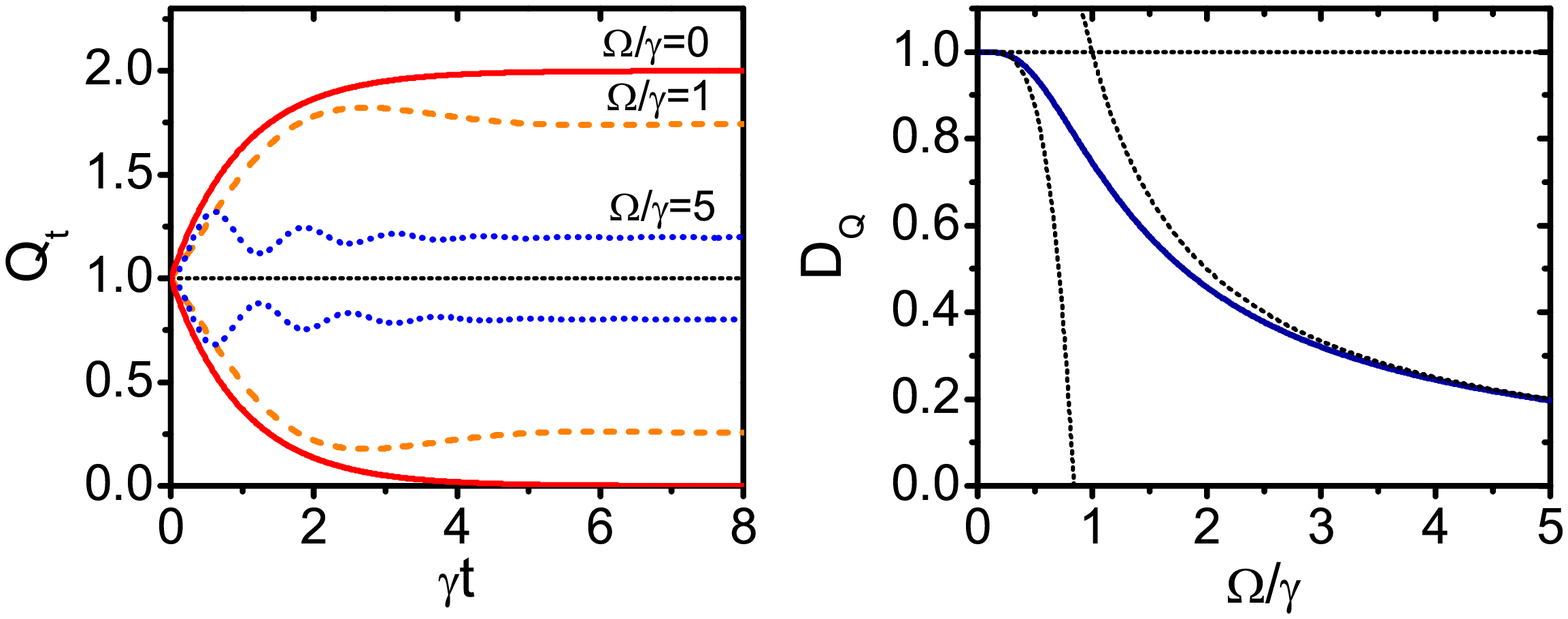}
\caption{Left panel, $Q_{t}$ [Eq.~(\protect\ref{QdeTe})] as a function of
time for different values of $\Omega /\protect\gamma .$ The system initial
conditions fulfill Eq.~(\protect\ref{Angles}). The curves above and below $%
Q_{t}=1$ correspond to the lower and upper initial states respectively.
Right panel, degree of quantumness $D_{Q}$ [Eq.~(\protect\ref{DQFluor})]
(full line) jointly the weak and strong intensity approximations [Eqs.~(%
\protect\ref{WeakAprox}) and (\protect\ref{DQRabiAlta})] (dotted lines).}
\end{figure}

With the previous election of initial conditions, from the definition~(\ref%
{Degree}) and Eq.~(\ref{QInfinito}), the degree of environment quantumness
associated to the dynamics~(\ref{Fluor}) can be written as 
\end{subequations}
\begin{equation}
D_{Q}=\frac{\gamma \sqrt{\gamma ^{2}+4\Omega ^{2}}}{\gamma ^{2}+2\Omega ^{2}}%
.  \label{DQFluor}
\end{equation}

In Fig.~1 (left panel) we plot the time-dependence of $Q_{t}$ [Eq.~(\ref%
{QdeTe})] assuming system initial conditions that maximize the degree of
environment quantumness [Eq.~(\ref{Angles})]. When $\Omega /\gamma =0,$ that
is in absence of the external excitation, it is obtained a monotonic
behavior, $Q_{t}=1\mp \lbrack 1-e^{-t\gamma }].$ Consistently, this case can
be recovered from Eq.~(\ref{QThermico}) after taking a vanishing environment
temperature $\langle \sigma _{z}\rangle _{\infty }=-1,$ and $\langle \sigma
_{z}\rangle _{0}=\pm 1.$ On the other hand, when increasing $\Omega /\gamma $
oscillations in the behavior of $Q_{t}$ emerges. In addition, the asymptotic
values $\lim_{t\rightarrow \infty }Q_{t}$ start to approach the unit value.
Even more, when $\Omega /\gamma \gg 1$ it follows that $\lim_{t\rightarrow
\infty }Q_{t}\approx 1.$

Given that for each value of $\Omega /\gamma $ the initial system state
fulfills the condition~(\ref{Angles}), the asymptotic values of $Q_{t}$
shown in Fig.~1 are related to the degree of environment quantumness [Eq.~(%
\ref{Degree})] as $D_{Q}=\left\vert \lim_{t\rightarrow \infty
}Q_{t}-1\right\vert .$ In the right panel of Fig.~1 we plot $D_{Q}$ as a
function of amplitude of the external field $\Omega /\gamma .$ When the
external excitation is weak, Eq.~(\ref{DQFluor}) can be well approximated by%
\begin{equation}
D_{Q}\simeq 1-2\left( \Omega /\gamma \right) ^{4}\ \ \ \ \ \ \ \ \left(
\Omega /\gamma \right) <1.  \label{WeakAprox}
\end{equation}%
Thus, in this regime the quantumness of the environment influence is
maximal. In fact, the departure from $D_{Q}=1$ depends on the fourth power
of $\Omega /\gamma .$ On the other hand, when the external excitation is
strong enough, it follows%
\begin{equation}
D_{Q}\simeq 1/(\Omega /\gamma )\rightarrow 0,\ \ \ \ \ \ \ \ \ \ \ (\Omega
/\gamma )\gg 1.  \label{DQRabiAlta}
\end{equation}%
This result means that, in this extreme regime, the environment influence
can be well approximated by classical noises. This is a non-intuitive
result. In fact, some quantum features of the dynamics~(\ref{Fluor}) emerge
when increasing the external coherent fields~\cite{carmichaelbook}. This
apparent contradiction is raised up when realizing that the proposed measure
quantifies how much the environment action departs from the influence of
classical noise fluctuations by considering the full open quantum dynamics,
that is, reservoir and external fields.

By finding the explicit solutions of the matrix elements of $\rho _{t},$
when $\Omega /\gamma \gg 1$ it is possible to approximate the Lindblad
evolution~(\ref{Fluor}) by%
\begin{equation}
\frac{d\rho _{t}}{dt}\approx -i\frac{\Omega }{2}[\sigma _{x},\rho _{t}]+%
\frac{3}{4}\gamma (\sigma _{z}\rho _{t}\sigma _{z}-\rho _{t}),\ \ \ \
(\Omega /\gamma )\gg 1.  \label{Dephasing}
\end{equation}%
Thus, the combined action of the environment (whose effective temperature is
cero) and the external excitation can be represented by a dephasing
mechanism, that is, $(3\gamma /4)(\sigma _{z}\rho _{t}\sigma _{z}-\rho
_{t}). $ This contribution can always be obtained by coupling the system to
external white noises such as for example Gaussian noises~\cite%
{GaussianNoise} or Poisson noises~\cite{viola}. Consequently, the
classicality indicated by the result~(\ref{DQRabiAlta}) is completely
consistent, which in turn also shows the physical meaning of the developed
approach.

\subsection{Optimal states for two-interacting qubits}

We consider two qubits $(a$ and $b)$ whose bipartite density matrix $\rho
_{t}^{ab}$\ evolves as%
\begin{equation}
\frac{d\rho _{t}^{ab}}{dt}=-i\frac{\Omega }{2}[\sigma _{x}\otimes \sigma
_{x},\rho _{t}^{ab}]+\gamma \mathcal{L}_{a}[\rho _{t}^{ab}]+\gamma \mathcal{L%
}_{b}[\rho _{t}^{ab}].  \label{TWO}
\end{equation}%
The frequency $\Omega $ scales the Hamiltonian interaction between both
systems. In addition, $\mathcal{L}_{a}$ and $\mathcal{L}_{b}$ define the
dissipative dynamics of each subsystem. They are defined by the dissipative
contribution in Eq.~(\ref{Fluor}), here written in each Hilbert space. We
study the relation between the proposed quantumness measure $Q_{t}$ and the
optimal initial conditions $\rho _{0}^{ab}$ that lead to its maximal value
in the stationary regime.

The stationary state $[\rho _{\infty }^{ab}=\lim_{t\rightarrow \infty }\rho
_{t}^{ab}]$ of the dynamics~(\ref{TWO}) can be obtained in an exact
analytical way. Introducing the standard base of states $\{|++\rangle
,|+-\rangle ,|-+\rangle ,|--\rangle \},$ it follows%
\begin{equation}
\rho _{\infty }^{ab}=\frac{1}{4\Gamma ^{2}}\left( 
\begin{array}{cccc}
\Omega ^{2} & 0 & 0 & -i2\gamma \Omega \\ 
0 & \Omega ^{2} & 0 & 0 \\ 
0 & 0 & \Omega ^{2} & 0 \\ 
i2\gamma \Omega & 0 & 0 & 4\gamma ^{2}+\Omega ^{2}%
\end{array}%
\right) ,
\end{equation}%
where $\Gamma \equiv \sqrt{\gamma ^{2}+\Omega ^{2}}.$ By using Eq.~(\ref%
{MaxEigen}), the degree of quantumness can be written in terms of the
largest eigenvalue of $\tilde{\rho}_{\infty }^{ab}=\rho _{\infty }^{\ast
ab}. $ We get%
\begin{equation}
D_{Q}=\frac{\gamma (\gamma +2\sqrt{\gamma ^{2}+\Omega ^{2}})}{\gamma
^{2}+\Omega ^{2}}.  \label{DQTWO}
\end{equation}%
In Fig.~2 (left panel) we plot $D_{Q}$ as a function of $\Omega /\gamma .$
We notice that by increasing the influence of the Hamiltonian contribution
classicality is achieved, $\lim_{\Omega /\gamma \rightarrow \infty }D_{Q}=0.$
Similarly to the previous case [Eqs.~(\ref{DQFluor}) and (\ref{Dephasing})],
in this limit the combined action of the environment and the subsystems
interaction Hamiltonian can be written in terms of dephasing mechanisms,
which in turn can be represented by the action of classical noise
fluctuations.%
\begin{figure}[tbp]
\includegraphics[bb=46 870 730 1132,angle=0,width=8.85cm]{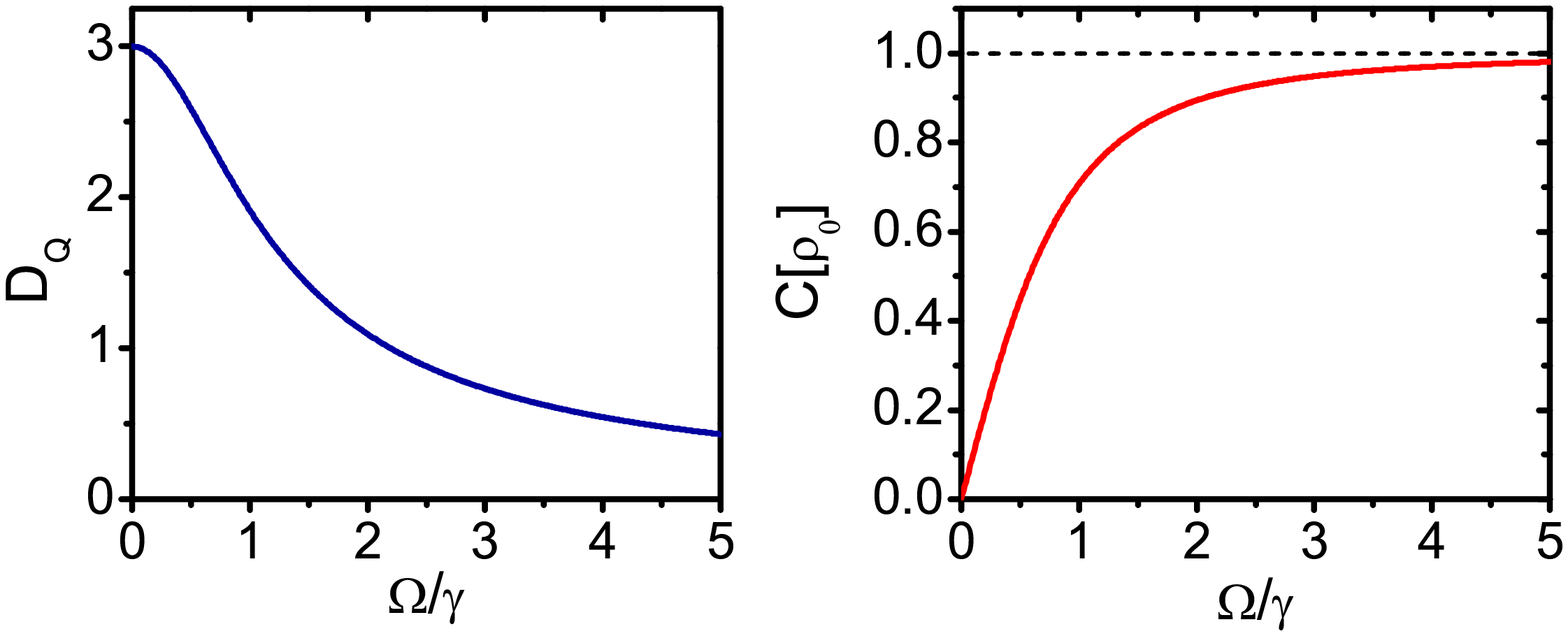}
\caption{Degree of environment quantumness $D_{Q}$ [Eq.~(\protect\ref{DQTWO}%
)] and concurrence of the initial optimal state $C[\protect\rho %
_{0}]=C[|i_{\max }\rangle \langle i_{\max }|]$ [Eq.~(\protect\ref{Imax})],
corresponding to the bipartite evolution~(\protect\ref{TWO}).}
\end{figure}

Eq.~(\ref{MaxEigen}) also characterize the initial condition $\rho
_{0}^{ab}=|i_{\max }\rangle \langle i_{\max }|$ that leads to maximal
stationary values of the quantumness measure $Q_{t}.$ $|i_{\max }\rangle $
is the eigenstate of the stationary state $\tilde{\rho}_{\infty }^{ab}$\
with the largest eigenvalue. It reads%
\begin{equation}
|i_{\max }\rangle =\frac{1}{\sqrt{2\Gamma (\Gamma -\gamma )}}[i(\Gamma
-\gamma )|++\rangle +\Omega |--\rangle ],  \label{Imax}
\end{equation}%
where as before $\Gamma =\sqrt{\gamma ^{2}+\Omega ^{2}},$ and $\langle
i_{\max }|i_{\max }\rangle =1.$

In general $|i_{\max }\rangle $\ is an entangled state. This feature can be
quantified through its concurrence~\cite{horodecki} $C[\rho _{0}]=C[|i_{\max
}\rangle \langle i_{\max }|].$ In Fig.~2 (right panel) we plot its
dependence with $\Omega /\gamma .$ In the limit of a vanishing unitary
coupling, from Eq.~(\ref{Imax}) it follows%
\begin{equation}
\lim_{\Omega /\gamma \rightarrow 0}|i_{\max }\rangle =|--\rangle .
\label{Cero}
\end{equation}%
This is an unentangled state, implying $C[\rho _{0}]=0.$ In contrast, in the
limit of strong coupling we get%
\begin{equation}
\lim_{\Omega /\gamma \rightarrow \infty }|i_{\max }\rangle =\frac{1}{\sqrt{2}%
}(i|++\rangle +|--\rangle ),  \label{Infinito}
\end{equation}%
which is a maximal entangled state, $C[\rho _{0}]=1.$

The previous behaviors have an interesting physical implication. In the weak
coupling limit $[\Omega /\gamma \approx 0],$ an (almost) unentangled initial
state leads to the maximal departure from classicality of the environment
action (quantified by $Q_{t}).$ When increasing the unitary coupling $%
[\Omega /\gamma >0],$ an increasing initial entanglement between both
subsystems is necessary to obtain the maximal departure from classicality.
In this way, entanglement becomes a necessary resource to detect the
quantumness of the environment influence when approaching a limit where a
classical noise approximation is valid.

For this model the quantumness measure $Q_{t}$ assumes a simple form [Eq.~(%
\ref{QDual})]. When maximizing its stationary value with respect to the
initial conditions it follows%
\begin{equation}
Q_{t}=1+\frac{\gamma ^{2}(1+e^{-2\gamma t})}{\Gamma ^{2}}+2\frac{\gamma }{%
\Gamma }\Big{[}1-\lambda e^{-2\gamma t}\cos (\Omega t)\Big{]},
\label{QBipartito}
\end{equation}%
where $\lambda \equiv 1+(\gamma /\Gamma ).$ Consistently, the initial state
that leads to this expression is $\rho _{0}^{ab}=|i_{\max }\rangle \langle
i_{\max }|$ [Eq.~(\ref{Imax})].

The previous results relies on taking both subsystems as the system of
interest. One can also deal with the partial dynamics $\rho _{t}^{a}=\mathrm{%
Tr}_{b}[\rho _{t}^{ab}],$ or alternatively $\rho _{t}^{b}=\mathrm{Tr}%
_{a}[\rho _{t}^{ab}].$ The corresponding stationary states read%
\begin{equation}
\rho _{\infty }^{s}=\frac{1}{2\Gamma ^{2}}\left( 
\begin{array}{cc}
\Omega ^{2} & 0 \\ 
0 & 2\gamma ^{2}+\Omega ^{2}%
\end{array}%
\right) ,\ \ \ \ \ \ s=a,b.
\end{equation}%
Performing similar calculations, the degree of quantumness and the optimal
state are%
\begin{equation}
D_{Q}=\frac{\gamma ^{2}}{\gamma ^{2}+\Omega ^{2}},\ \ \ \ \ \ \ \ |i_{\max
}\rangle =|-\rangle .
\end{equation}%
Given the symmetry of Eq. (\ref{TWO}), this results applies to both
subsystems. Furthermore, assuming $\rho _{0}^{ab}=|i_{\max }\rangle \langle
i_{\max }|\otimes \rho _{0}^{b},$ where $\rho _{0}^{b}$ is an arbitrary
state, it follows%
\begin{equation}
Q_{t}=1+\frac{\gamma ^{2}}{\Gamma ^{2}}+\frac{\gamma e^{-\gamma t}}{\Gamma
^{2}}\Big{[}\Omega \sin (\Omega t)-\gamma \cos (\Omega t)\Big{]}.
\end{equation}%
The same expression follows from $\rho _{0}^{ab}=\rho _{0}^{a}\otimes
|i_{\max }\rangle \langle i_{\max }|.$ These results differ from those
obtained starting from a bipartite representation [Eqs.~(\ref{DQTWO}), (\ref%
{Imax}) and~(\ref{QBipartito})]. This feature shows that the environment
influence over a system cannot in general be related in a simple way with
the action over the constitutive subsystems.

\subsection{Quantum harmonic oscillator coupled to a thermal environment}

The developed approach applies consistently to systems with a Hilbert space
of finite dimension [see Eqs.~(\ref{Bounds}) and~(\ref{DqBounds})].
Complementarily, here we study the case of a quantum harmonic oscillator
coupled to a thermal environment at a finite temperature.

The density matrix evolution can be written as in Eq.~(\ref{Thermal}) under
the replacements $\sigma ^{\dagger }\rightarrow a^{\dagger }$ and $\sigma
\rightarrow a,$ where $a^{\dag }$ and $a$ are the creation and annihilation
Bosonic operators of the system respectively~\cite{breuerbook}. The
evolution can alternatively be written through a Wigner function. It is
defined as the Fourier transform $W(\alpha ,\alpha ^{\ast },t)\equiv (1/\pi
^{2})\int d^{2}z\chi (z,z^{\ast })e^{-iz^{\ast }\alpha ^{\ast }}e^{-iz\alpha
},$ where the characteristic function is$\ \chi (z,z^{\ast })\equiv \mathrm{%
Tr}_{s}[\rho _{t}\exp (iz^{\ast }a^{\dag }+iza)].$ Denoting $W_{t}=W(\alpha
,\alpha ^{\ast },t),$ its time evolution reads~\cite{carmichaelbook}%
\begin{equation}
\frac{\partial W_{t}}{\partial t}=\Big{\{}\varphi \frac{\partial }{\partial
\alpha }\alpha +\varphi ^{\ast }\frac{\partial }{\partial \alpha ^{\ast }}%
\alpha ^{\ast }+\Big{(}\frac{\kappa +\zeta }{2}\Big{)}\frac{\partial ^{2}}{%
\partial \alpha \partial \alpha ^{\ast }}\Big{\}}W_{t},  \label{Wigner}
\end{equation}%
where $\varphi \equiv i\omega _{0}+(\kappa -\zeta )/2.$ Here, $\omega _{0}$
is the natural frequency of the system. Notice that dissipative
contributions (first-order derivatives) are present whenever the underlying
rates are different, $\kappa \neq \zeta .$ On the other hand, diffusion
(second-order derivatives) always develops, being scaled by $(\kappa +\zeta
)/2.$

The operator evolution can be obtained in a similar way from the dual
dynamics associated to the system density matrix. Alternatively, it can be
deduced by using that operator expectation values can be written as $\langle
A\rangle _{t}=\int d\alpha d\alpha ^{\ast }W_{t}A_{0}(\alpha ,\alpha ^{\ast
})=\int d\alpha d\alpha ^{\ast }W_{0}A(\alpha ,\alpha ^{\ast },t),$ where $%
A_{0}(\alpha ,\alpha ^{\ast })$ is the \textquotedblleft scalar
representation\textquotedblright\ of the system operator $A.$ From Eq.~(\ref%
{Wigner}) we get $[A_{t}=A(\alpha ,\alpha ^{\ast },t)]$%
\begin{equation}
\frac{\partial A_{t}}{\partial t}=-\Big{\{}\varphi \alpha \frac{\partial }{%
\partial \alpha }+(\varphi \alpha )^{\ast }\frac{\partial }{\partial \alpha
^{\ast }}-\Big{(}\frac{\kappa +\zeta }{2}\Big{)}\frac{\partial ^{2}}{%
\partial \alpha \partial \alpha ^{\ast }}\Big{\}}A_{t}.  \label{DualWigner}
\end{equation}%
Using this representation, the quantumness indicator [Eq.~(\ref{QDual})] can
be expressed as $Q_{t}=\int d\alpha d\alpha ^{\ast }A_{t},$ where $A_{t}$ is
the solution of the dual evolution with initial condition $A_{0}=W_{0}.$ By
integration by parts of Eq.~(\ref{DualWigner}) it is simple to arrive to $%
(d/dt)Q_{t}=(\kappa -\zeta )Q_{t},$ which leads to%
\begin{equation}
Q_{t}=\exp [(\kappa -\zeta )t].  \label{QHarmonico}
\end{equation}%
This results is valid independently of the initial system density matrix.
Thus, not any maximization procedure is available. The same expression for $%
Q_{t}$ follows by using a Glauber-Sudarshan P-representation or
Q-representation~\cite{carmichaelbook}, or even from a (diagonal)
Fock-number characteristic function approach~\cite{orzag}.

Given that $\kappa \geq \zeta $ $[\kappa =\gamma (n_{th}+1)$ and $\zeta
=\gamma n_{th}],$ the indicator $Q_{t}$ develops an exponential divergence
in time for any finite temperature of the bath. Only when the reservoir
temperature is infinite $(\kappa =\zeta )$ a classical noise representation
applies, $Q_{t}=1$ and $D_{Q}=0.$ This behavior has a clear interpretation.
In fact, for any finite reservoir temperature, the Wigner function involves
dissipative contributions [see Eq.~(\ref{Wigner})]. These (trace preserving)
effects develop in the system Hilbert space and cannot be reproduced by any
classical external influence. Dissipative contributions only vanishes when $%
\kappa =\zeta ,$ which consistently supports the (discontinuous)
temperature-dependence of the degree of quantumness in this case.

While the previous result is consistent it strongly differs from the
two-level system case [see Eq.~(\ref{DqThermal})], where $D_{Q}$ has a
continuous dependence on the reservoir temperature. Interestingly, for
Hilbert spaces of infinite dimension the expression~(\ref{DqInfinita})
allows us to define a \textit{renormalized degree of quantumness} as $%
D_{Q_{R}}\equiv \max_{\lbrack \rho _{0}]}\mathrm{Tr}_{s}[\tilde{\rho}%
_{\infty }\rho _{0}].$ In terms of the Wigner function it reads%
\begin{equation}
D_{Q_{R}}=\max_{[W_{0}]}\int d\alpha d\alpha ^{\ast }\tilde{W}_{\infty
}W_{0}.  \label{DQWigner}
\end{equation}%
Here, the maximization must be performed over all possible (normalized)
initial conditions $W_{0}.$ In addition, Eq.~(\ref{Wigner}) implies that $%
\tilde{W}_{\infty }=W_{\infty }=\lim_{t\rightarrow \infty }W_{t}=(1/\pi
\sigma _{\infty })\exp (-|\alpha |^{2}/\sigma _{\infty }),$ where $\sigma
_{\infty }=(1/2)(\kappa +\zeta )/(\kappa -\zeta )=n_{th}+(1/2)=(1/2)(\tanh
[\beta \hbar \omega _{0}/2])^{-1}.$

Given that $D_{Q_{R}}$ is a linear functional of $W_{0}$ the maximization
problem cannot be solved by using standard functional derivative techniques.
As an ansatz we assume that $W_{0}$ is also a Gaussian function. In such a
case, it follows that $W_{0}$ must has the minimal possible wide. Thus, it
must be the Wigner function of the ground state of the system, which in turn
from Eq.~(\ref{DQWigner}) delivers%
\begin{equation}
D_{Q_{R}}=\frac{1}{\sigma _{\infty }+(1/2)}=1-\exp (-\beta \hbar \omega
_{0}).
\end{equation}%
The same result follows by performing a similar ansatz in the energy
eigenbasis representation. $D_{Q_{R}}$ has the expected dependence with the
environment temperature. In particular, classicality $[D_{Q_{R}}=0]$ is
approached in a high temperature limit. In contrast to the two-level system
[Eq.~(\ref{DqThermal})], here the renormalized degree of quantumness cannot
be associated to the time-dependent quantumness measure $Q_{t}$ [Eq.~(\ref%
{QHarmonico})].

\section{Summary and Conclusions}

We have developed a consistent proposal that allows quantifying how far the
influence of a given environment over an open quantum system departs from
the action of classical stochastic fields. Its physical ground relies on
associating the quantumness of the environment influence with the lack of
commutativity between the reservoir initial state and the total
system-environment Hamiltonian. Over this basis we introduced a
(time-dependent) quantumness measure [Eq.~(\ref{QDef})]. Its stationary
value (long time-regime) when maximized over all possible system initial
conditions define a degree of environment non-classicality [Eq.~(\ref{Degree}%
)]. For dissipative dynamics it can be determine from the largest eigenvalue
of the (time reversal) stationary system density matrix [Eq.~(\ref{MaxEigen}%
)]. Independently of the system dynamical regime, the quantum measure can be
written in terms of the operators dual evolution [Eq.~(\ref{QDual})]. This
alternative definition provides a powerful tool for characterizing the
quantumness measure in both Markovian and non-Markovian regimes.

Consistently the quantumness measure vanishes identically for a wide class
of quantum dynamics, which include Hamiltonian ensembles [Eq.~(\ref%
{HEnsemble})], stochastic Hamiltonians [Eq.~(\ref{Ruido})], and a class of
collisional dynamics [Eq.~(\ref{UnitalCollision})]. 
All of these dynamics can be obtained by considering the action of
underlying classical stochastic processes. In spite of the consistence of
this result, the quantumness indicator also vanishes when the open system
dynamics is defined by a unital map [Eq.~(\ref{Unital})]. Hence, the
proposed indicator can also be read as a measure of departure from this
dynamical property.

The consistence of the developed approach was supported by studying
different dissipative open system dynamics. For two-level systems coupled to
a thermal bath, the degree of environment quantumness decreases
monotonically with the reservoir temperature [Eq.~(\ref{DqThermal})]. For an
optical transition (resonant fluorescence) the amplitude of the external
coherent excitation monotonically drives the environment influence to
classicality [Eq.~(\ref{DQFluor})]. The consistence of this result follows
from the possibility of describing the high intensity regime in terms of a
dephasing quantum master equation that can be represented by the action of
classical noises. On the other hand, by analyzing two interacting qubits it
was found that quantum entanglement may become a necessary resource for
detecting the quantumness of the environment influence when approaching a
regime where a classical noise representation becomes a valid approximation.
Application to systems endowed with a Hilbert space of infinite dimension
was also established.

The present formalism lefts open some interesting issues. For example, which
dynamical features determine the presence or absence of revivals in the
time-behavior of the quantumness indicator is unknown. On the other hand,
and operational definition and experimental measurability are also
interesting issues that can be tackled from the proposed approach.

\section*{Acknowledgments}

Valuable discussions with Oscar Mensio are gratefully acknowledged. A.A.B.
also thanks support from Consejo Nacional de Investigaciones Cient\'{\i}%
ficas y T\'{e}cnicas (CONICET), Argentina.

\appendix*

\section{Renewal collisional models}

In these models the statistics of the collisional times are defined by a
\textquotedblleft waiting time distribution\textquotedblright\ $w(t).$ It
gives the probability density for the time interval between consecutive
collisional events. The corresponding survival probability is defined as $%
P_{0}(t)=1-\int_{0}^{t}dt^{\prime }w(t^{\prime }).$ Poisson statistics
corresponds to $w(t)=\gamma \exp (-\gamma t),$ $P_{0}(t)=\exp (-\gamma t),$
assumption that lead to Markovian Lindblad equations for the system dynamics.

In correspondence with Eq.~(\ref{Colision}), the system density matrix can
be written in general as~\cite{collisional}%
\begin{equation}
\rho _{t}=\sum_{n=0}^{\infty }\int_{0}^{t}dt^{\prime }\mathcal{P}%
_{0}(t-t^{\prime })\mathcal{W}^{(n)}(t^{\prime })\rho _{0}.  \label{Average}
\end{equation}%
The involved superoperators are written in a Laplace domain $%
[f(u)=\int_{0}^{\infty }dte^{-ut}f(t)]$ as%
\begin{equation}
\mathcal{P}_{0}(u)\equiv P_{0}(u-\mathcal{L}_{s}),\ \ \ \ \ \ \mathcal{W}%
^{(n)}(u)\equiv \lbrack \mathcal{E}w(u-\mathcal{L}_{s})]^{n}.
\label{Woperator}
\end{equation}%
Here, $P_{0}(u)=[1-w(u)]/u.$ Furthermore, the free propagator between events
was written as $\mathcal{G}_{t}=\exp (t\mathcal{L}_{s}).$ Notice that in a
time domain $\mathcal{W}^{(n)}(t)$ consists in the convolution of free
propagation and $n$-collisional events. Consistently, the function $%
P_{0}(u)w^{n}(u)$ gives the probability of occurring $n$-events up to time $%
t.$

Using that $\langle A\rangle _{t}=\mathrm{Tr}_{s}[A_{0}\rho _{t}]=\mathrm{Tr}%
_{s}[\rho _{0}A_{t}],$ from Eq.~(\ref{Average}) the operator dual evolution
reads%
\begin{equation}
A_{t}=\sum_{n=0}^{\infty }\int_{0}^{t}dt^{\prime }\mathcal{W}^{\bigstar
(n)}(t^{\prime })\mathcal{P}_{0}^{\bigstar }(t-t^{\prime })A_{0}.
\label{DualAverageA}
\end{equation}%
Here, the involved superoperators are defined as $\mathcal{P}_{0}^{\bigstar
}(z)=P_{0}(z-\mathcal{L}_{s}^{\bigstar })$ and $\mathcal{W}^{\bigstar
(n)}(z)=[w(z-\mathcal{L}_{s}^{\bigstar })\mathcal{E}^{\bigstar }]^{n}.$
These expressions rely on the definitions $\mathrm{Tr}_{s}[A\mathcal{E}[\rho
]]=\mathrm{Tr}_{s}[\rho \mathcal{E}^{\bigstar }[A]],$ and $\mathrm{Tr}%
_{s}[A\exp (t\mathcal{L}_{s})[\rho ]]=$ $\mathrm{Tr}_{s}[\rho \exp (t%
\mathcal{L}_{s}^{\bigstar })[A]].$

The time-dependent quantumness indicator $Q_{t}$ can be calculated as the
trace of dual dynamics [Eq.~(\ref{QDual})]. The property $\mathrm{Tr}%
_{s}[A_{t}]=\mathrm{Tr}_{s}[A_{0}]$ leads to the condition $\mathrm{Tr}_{s}[%
\mathcal{E}^{\bigstar }[A]]=\mathrm{Tr}_{s}[A],$ which recovers Eq.~(\ref%
{UnitalCollision}).

\end{document}